\documentstyle[prl,aps,epsf]{revtex}

\newcommand{\TKT}{T_{\rm KT}}

\newcommand{\vort}{{\mathrm vort}}
\newcommand{\Eq}[1]{Eq.~(\ref{#1})}
\newcommand{\Tkt}{0.765}

\begin{document}
\advance\textheight by 0.2in
\draft
\twocolumn[\hsize\textwidth\columnwidth\hsize\csname@twocolumnfalse\endcsname

\title{The Transition in the Two-Dimensional Step Model: A
  Kosterlitz-Thouless Transition in Disguise}

\author{Peter Olsson and Petter Holme}

\address{Department of Theoretical Physics, Ume{\aa} University, 
  901 87 Ume{\aa}, Sweden}

\date{\today}   

\maketitle

\begin{abstract}
  Evidence for a Kosterlitz-Thouless transition in the 2D step model
  is obtained from Monte Carlo determinations of the helicity modulus.
  It is argued that the free energy of a single vortex at the center
  of the system depends logarithmically on the system size in spite of
  the fact that the spin interaction is not harmonic for small
  differences in the spin angles. We conclude that this is the reason
  for the Kosterlitz-Thouless transition in the 2D step model and that
  the harmonic spin interaction not is a necessary requirement.
\end{abstract}

]

The phase transition in two-dimensional (2D) XY models is known to
take place through the vortex unbinding mechanism due to Kosterlitz and
Thouless (KT)\cite{Kosterlitz_Thouless}. From the principles of
universality one expects this transition to remain the same
independent of details of the system as e.g.\ the underlying lattice
structure. The precise spin interaction potential $U(\phi)$, where
$\phi$ is the angle difference between neighboring spins, is not
supposed to be essential either.  $U(\phi)$ is however required to be
periodic in $2\pi$ and it seems always to have been presumed that the
interaction in addition has to be harmonic for small $\phi$.  The
harmonicity for small $\phi$ has to do with the energetics for vortex
formation.  With a harmonic potential the energy for a single vortex
in a $L\times L$ lattice goes as $\ln L$, and in the classical
argument by Kosterlitz and Thouless this property is crucial for the
transition.

The subject of the present Letter is the 2D step model which is an XY
model with a spin interaction that has no harmonic component.
$U(\phi)$ is instead a step-like function
\begin{equation}
  U(\phi) = -J\, {\rm sign}(\cos\phi)
  \label{Uphi}
\end{equation}
Since this potential is flat around $\phi = 0$ there is no $\ln
L$-dependence of the energy for a single vortex. This energy is
instead independent of system size and one would therefore expect a
non-vanishing density of free vortices at all finite temperatures, and
consequently no phase transition.  Against that background the
evidence from simulations for a transition were very intriguing.  The
first clear evidence for a transition was obtained from a Monte Carlo
(MC) study of the susceptibility and the specific
heat\cite{Sanchez-Velasco_Wills}. The increase of the susceptibility
with lattice size was considered to suggest a phase transition at $T
\approx 1.1$. Later simulations also provided evidence that the
transition actually is in the same universality class as the harmonic
XY models\cite{Irving.Kenna}.  The similarity of the behavior close to
the transition of an harmonic XY model and the step model also led
these authors to question the vortex unbinding as the mechanism behind
the KT transition\cite{Irving.Kenna}.  How could a transition driven
by vortices be altogether insensitive to the very different energy
cost for vortices in the two models?

In the present Letter we address the question of the necessity of an
harmonic spin potential for the KT transition by examining the
behavior of the 2D step model. We first calculate the helicity modulus
and show that the behavior of this quantity gives strong support for a
KT transition.  We then demonstrate that the cost in \emph{free
  energy} for a single vortex at the center of the system in fact goes
as $\ln L$.  It is this feature that stabilizes the low temperature
phase.  Finally, we refine the arguments to obtain quantitatively
satisfactory estimates.

The helicity modulus $\Upsilon$, is a convenient quantity for the
study of KT transitions due to its universal value $2T/\pi$ at the
transition\cite{Nelson_Kosterlitz,Minnhagen_Warren}, and the known
form of the approach to this universal value with
$L$\cite{Weber_Minnhagen:88}. The usual procedure in MC simulations is
to determine $\Upsilon$ from a correlation function which involves
some derivatives of $U(\phi)$\cite{Ohta_Jasnow}.  Clearly, with a
step-like potential the derivatives of the potential cannot be
calculated and this expression cannot be used. A way out is to instead
start from the defining expression for the helicity modulus
\begin{equation}
  \Upsilon = \left.\frac{\partial^2 F}{\partial
  \Delta^2}\right|_{\Delta = 0},
  \label{UpsF}
\end{equation}
and perform the simulations with fluctuating twist boundary
conditions\cite{Olsson.twist}. In these simulations one collects a
histogram of the total twist $P(\Delta)$.  Since the probability for a
certain twist is related to the free energy through $P(\Delta) \propto
e^{-F(\Delta)/T}$ \Eq{UpsF} becomes
\begin{equation}
  \Upsilon = -T \left.\frac{\partial^2 \ln P}{\partial
  \Delta^2}\right|_{\Delta = 0}.
  \label{UpsP}
\end{equation}
The simulations are done with twist variables in the two directions,
$\Delta_x$ and $\Delta_y$, which beside the spin variables $\theta_i$
are updated with the Metropolis algorithm.  With ${\mathbf r}_{ij}$ a
unit vector between nearest neighbors and ${\mathbf\Delta} =
(\Delta_x,\Delta_y)$, the Hamiltonian may be written
\begin{displaymath}
  H = \sum_{\langle ij\rangle} U\left(\theta_i - \theta_j -
  \frac{1}{L}{\mathbf r}_{ij} \cdot 
  {\mathbf\Delta}\right) = \sum_{\langle ij\rangle} U(\phi_{ij}).
\end{displaymath}

To get a good acceptance ratio for the twist update it is necessary to
make use of $L$ different twist variables in each direction, with
$\Delta_x = \sum_{k=1}^L \Delta_x^{(k)}$, (and similarly in the $y$
direction) where $k$ is the column (row) number. In our simulations,
which for convenience were for a $O(256)$ model, we used the potential
$U(\phi) = -J$ for 129 angle differences $[-\pi/2$, $\pi/2]$ and $+J$
for the remaining 127\cite{angle129}. This choice is not expected to
be important for the transition properties, but gives a slight shift
of the transition temperature as compared to the potential of
\Eq{Uphi}. The length of the runs were typically $5\times 10^8/L$
sweeps through the lattice.

\begin{figure}
  \epsfxsize=8.5truecm
  \epsffile[105 320 609 580]{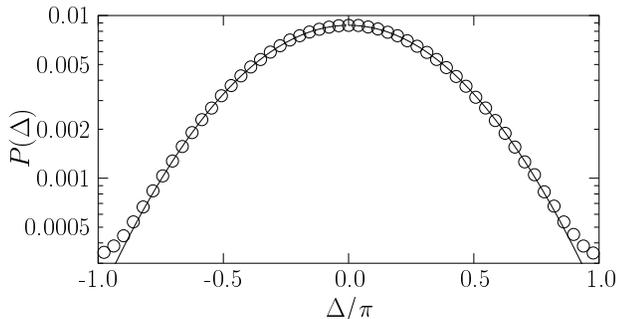}
  \caption{Twist histogram obtained through MC simulations at
    $T=0.05$. The solid line is obtained by fitting the data to a
    quadratic form which gives $\Upsilon/T = 0.789$.}
  \label{fig-lnP.Delta}
\end{figure}

In Fig.\ \ref{fig-lnP.Delta} we show the histogram $P(\Delta)$ from
Monte Carlo simulations at $T = 0.05J$. Since the histogram is peaked
around $\Delta = 0$ the figure immediately gives evidence for a low
temperature phase with a finite stiffness. To determine $\Upsilon$ we
fit a quadratic curve to $\ln P$ for $|\Delta/\pi|< 1/3$ and obtain
$\Upsilon/T = 0.789(4)$ (where the given error is one standard
deviation). Note that this is slightly larger than the universal value
$2/\pi \approx 0.637$, which is required for a stable low temperature
phase.
An important feature of the step model is the gap in excitation
energies; there are no excitations with energy $< 2J$. At $T \ll J$
the system is therefore at all times in one of its numerous ground
states which means that the histogram $P(\Delta)$ is independent of
temperature.  From \Eq{UpsP} then follows a linear temperature
dependence for $\Upsilon$.  This is in contrast to harmonic XY models
for which $\Upsilon/J$ approaches unity in the low-temperature limit.

\begin{figure}
  \epsfxsize=8.5truecm
  \epsffile[105 320 609 660]{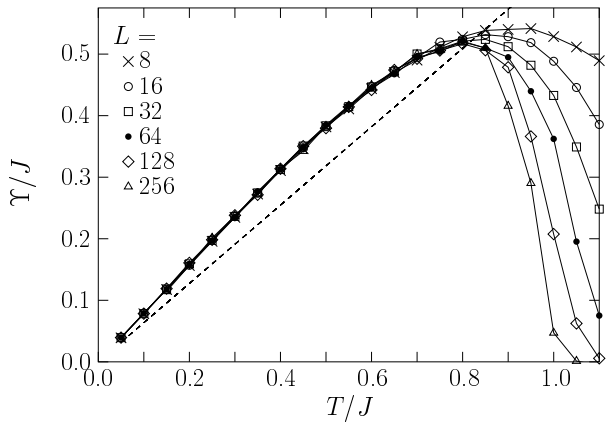}
  \caption{The helicity modulus versus temperature for $L = 8$ through
    256. The rapid decrease down to zero directly follows the crossing
    of the data with the universal line, $2T/\pi$ (dashed line). This
    behavior is suggestive of a KT transition. Note that even though
    $\Upsilon\rightarrow 0$ at low temperatures, the system remains in
    the low temperature phase since $\Upsilon/T > 2/\pi$.}
  \label{fig-Ups.T}
\end{figure}

The temperature dependence of $\Upsilon$ is shown in
Fig.~\ref{fig-Ups.T} for several system sizes together with the dashed
line for the universal jump condition $2T/\pi$.  We note that the
curves start out linearly at low temperatures, become size-dependent
at $T/J \approx 0.75$, cross the universal line and then drop down to
zero.  Beside the unusual linear temperature-dependence at low $T$
this behavior is just as in an ordinary harmonic XY model and
therefore precisely what one would expect for a KT transition.

\begin{figure}
  \epsfxsize=8.5truecm
  \epsffile[105 320 609 660]{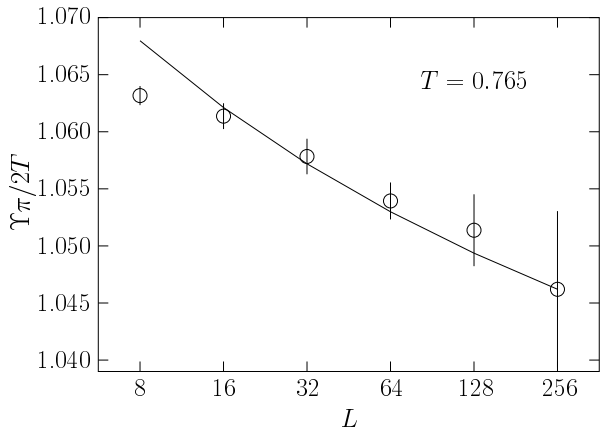}
  \caption{Determination of the transition temperature. The solid line
  is \Eq{eq-weber} and the data points are the helicity modulus from
  our simulations.}
  \label{fig-Ups.L}
\end{figure}

To determine the KT temperature we make use of the finite size
dependence of $\Upsilon$\cite{Weber_Minnhagen:88}. We follow the
procedure in Ref.\ \cite{Olsson:Kost-fit} of first fitting our MC data
for $\Upsilon$ from a narrow temperature interval to second order
polynomials in $T$, one for each $L$, and then fit the data to the
expression

\begin{equation}
  \Upsilon_L(\TKT) = \frac{2\TKT}{\pi}\left(1 + \frac{1}{A + 2\ln
  L}\right).
  \label{eq-weber}
\end{equation}
which amounts to adjusting $\TKT$ and $A$ to get the best possible
fit.  Using data for $L\geq 16$ we obtained $\TKT = \Tkt(6)$.  Fig.\ 
\ref{fig-Ups.L} illustrates the good fit of the data at the transition
temperature to the line from \Eq{eq-weber}. We consider the above MC
data to be strong evidence for a KT transition. This is in agreement
with the conclusion in Ref.~\cite{Irving.Kenna} that the step model is
in the same universality class as the harmonic XY model.

We now propose an analytical analysis in order to understand this
behavior.  We focus on the properties in the low temperature phase
where the angular differences are restricted to the low energy region,
$|\phi| \leq \pi/2$.  A central idea in the present letter is to note
that, while in the harmonic XY model the spinwave-vortex interaction
only is a smooth perturbation, in the step model this interaction
leads to a dramatic and crucial effect: while the energy of a system
with a single vortex fixed in the center of the system is finite, the
\emph{free energy} of this system grows as $\ln L$. This is due to the
change in entropy of spinwave fluctuations for the configuration with
the fixed vortex, as compared to the vortex free case. Note that this
is the entropy associated with a \emph{fixed} vortex, not the
positional entropy associated with a free vortex's variable location.

\begin{figure}
  \epsffile[0 -15 100 95]{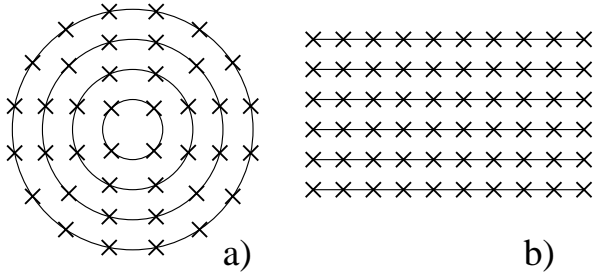}
  \caption{Arrangement of links used to a) argue for the change in
    spinwave entropy for a vortex at a fixed position and b) estimate
    the helicity modulus.}
  \label{fig-circles}
\end{figure}

To demonstrate the existence of this spinwave entropy we consider the
configuration of spins in Fig.\ \ref{fig-circles}a where we slightly
reorganize the spins and delete the links in the radial direction.  We
will return below to the approximation involved in this step.

The condition for having a positive vortex in the center of the system
in Fig.\ \ref{fig-circles}a is that the phase rotates by $2\pi$ along
each of the circles.  We therefore introduce $\Phi_r = \sum\phi$ along
the circle with radius $r$, which is a sum of $\approx 2\pi r$ values
and let $\Omega_r(\Phi_r)$ denote the number of possible combinations
of the $\phi$:s at distance $r$ as a function of $\Phi_r$, which is
defined only for $\Phi_r = 2\pi n$ (with integer $n$). The probability
for having a (positive) vortex is then determined by the product
\begin{equation}
  P_\vort \approx \prod_{r=1}^L\frac{\Omega_r(2\pi)}{\Omega_r(0)}
  \label{Pvort}
\end{equation}
The fraction $\Omega_r(2\pi)/\Omega_r(0)$ may be calculated if we
temporarily open up a closed path that makes up a circle. Since we
want the same number of links along this path we need one more spin
variable at one of the endpoints, which we take to be independent of
the other endpoint.  $\Phi_r$ then becomes a sum of $\approx 2\pi r$
independent variables $\phi$.  With the average of $\phi$ being equal
to zero and its variance given by $\sigma^2$ the distribution of
$\Phi_r$ for this open path becomes a Gaussian with width $2\pi
r\sigma^2$:
\begin{equation}
  \Omega_r^{\mathrm open}(\Phi_r) \propto
  \exp\left(-\frac{\Phi_r^2}{2\pi r\times2\sigma^2}\right)
  \label{Oopen}
\end{equation}
We now make use of the fact that the number of possible configurations
for the open and closed paths are the same if the spins at the
endpoints of the open path are equal.  Since this condition is equal
to having $\Phi_r = 2\pi n$ we conclude that $\Omega_r(2\pi n) =
\Omega_r^{\mathrm open}(2\pi n)$. From Eqs.~(\ref{Pvort}) and
(\ref{Oopen}) the probability for a vortex then becomes
\begin{displaymath}
  P_\vort
  = \exp\left(-\frac{\pi}{\sigma^2}\sum_{r=1}^L\frac{1}{r}\right)
  \approx \exp\left(-\frac{\pi}{\sigma^2}\ln L\right),
\end{displaymath}
for the entropy of a fixed vortex we obtain $S_\vort = \ln P_\vort =
-\frac{\pi}{\sigma^2}\ln L$, and the free energy for a vortex at a
fixed position in a system of size $L$ finally becomes
\begin{equation}
  F_\vort = -T S_\vort = T \frac{\pi}{\sigma^2}\ln L.
  \label{Fvort}
\end{equation}

For the harmonic XY model the well-known argument for the phase
transition gives the free energy for having a vortex at any of the
$L^2$ positions as
\begin{displaymath}
  \Delta F = \left(\pi J - 2T\right) \ln L,
\end{displaymath}
and the transition takes place at $T/J = 2/\pi$. In the step model the
corresponding expression becomes
\begin{displaymath}
  \Delta F = F_\vort - 2 T\ln L = T \left(\frac{\pi}{\sigma^2} -
  2\right) \ln L,
\end{displaymath}
where the temperature, at first sight only appears as a prefactor.
However, there is a hidden temperature dependence in the variance
$\sigma^2$. At low enough temperatures the $\phi_{ij}$ are restricted
to the interval $[-\pi/2,\pi/2]$ but with increasing temperature the
$\phi_{ij}$ will more often take values outside this interval, and
$\sigma^2$ will increase.  Therefore, if $\Delta F$ is positive at low
temperatures it will turn negative at some finite temperature and this
will give the transition. But if $\sigma^2 > \pi/2$ already at low
temperatures there will be no transition.

To see how the local restrictions on the angle differences give rise
to the non-zero helicity modulus we now turn to a rectangular geometry
and consider the determination of $\Upsilon$ from the distribution of
$\Delta_x$ and $\Delta_y$.  One point with examining the distribution
of the twist is to give predictions that are easy to compare with MC
simulations.  In the simplest approximation we again delete all links
in the perpendicular direction, as in Fig.~\ref{fig-circles}b. For a
single row the number of configurations consistent with a certain
total twist becomes $\Omega_{\mathrm row}(\Delta) \propto
\exp\left(-\Delta^2/2 L\sigma^2\right)$, in analogy with \Eq{Oopen},
and the number of configurations for the whole system with $L$ rows
becomes
\begin{displaymath}
  \Omega(\Delta) = \left[\Omega_{\mathrm row}(\Delta)\right]^L \propto
%  \exp\left(-\frac{\Delta^2}{2 \sigma^2}\right).
  \exp\left(-\Delta^2/2 \sigma^2\right).
\end{displaymath}
For the free energy we then arrive at $F(\Delta) =
T\Delta^2/2\sigma^2$ and with \Eq{UpsF} the helicity modulus becomes
\begin{equation}
  \Upsilon = T/\sigma^2.
  \label{Ups-sigma}
\end{equation}
In the absence of perpendicular links and at low temperatures, the
$\phi_{ij}$ have a rectangular distribution, and from elementary
integrals one finds $\sigma^2 = \pi^2/12 \approx 0.822$.  Through
\Eq{Ups-sigma} this gives $\Upsilon/T \approx 1.22$ which is about
50\% larger than $\Upsilon/T \approx 0.789$ from Monte Carlo
simulations, cf.\ Fig.~\ref{fig-lnP.Delta}. A better estimate will be
obtained below by including some links in the perpendicular direction.

Comparing Eqs.~(\ref{Fvort}) and (\ref{Ups-sigma}) we find $F_\vort(L)
= \Upsilon\pi\ln L$ which is the same relation as in the harmonic
model.  This shows that our two different calculations are equivalent
which is a consequence of using the same approximation in both cases,
i.e.\ neglecting all links perpendicular to the direction of interest.

We now discuss the assumption used above, namely that the
qualitatively correct behavior in a certain direction may be obtained
even though one neglects the perpendicular links.  This assumption is
true only if the relative reduction of the number of allowed
configurations obtained by introducing perpendicular links is
essentially independent of the total twist in the direction of
interest.  We argue that this is a plausible assumption by considering
two sets of configurations: 1) the set of all twist free
configurations and 2) the set of configurations with a twist
$\Delta_x$. There is then a transformation $\phi_{ij} + \Delta/L
\rightarrow \phi_{ij}$ on all the horizontal links that transforms
each member in the twist free set into a corresponding one in the
twisted set. Since this transformation not affects the angle
difference at the perpendicular links, it follows that the effect of
the perpendicular links will be to exclude the same number of
configurations in both these sets and this suggests that the relative
reduction due to the perpendicular links will be independent of
$\Delta$. However, this argument only serves to make our assumption a
reasonable one; it is not conclusive. There is nothing that guarantees
that the relative reduction of the \emph{allowed} configurations (with
$|\phi_{ij}| < \pi/2$ for all horizontal links) will be the same for
the two different sets. That a certain member of the twist free set is
allowed doesn't imply that the corresponding member in the twisted set
is allowed too.

\begin{figure}
  \epsffile[0 0 230 86]{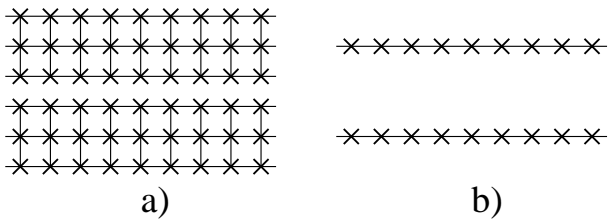}
  \caption{The configuration of links used to get quantitatively
    reasonable values. The starting point is given in panel a) and
    after integrating out the upper and lower rows on each triple row
    one obtains the links shown in panel b).}
  \label{fig-three-rows}
\end{figure}

\begin{figure}
  \epsffile[160 531 284 554]{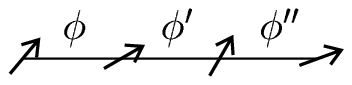}
  \caption{Illustration to the link-link correlations in
    \Eq{sigma-eff}.}
  \label{fig-prim-bis}
\end{figure}

Even though we have argued that the perpendicular links may be
neglected in a qualitative discussion, they have to be included in
order to get reasonable quantitative estimates, since they do affect
the variance $\sigma^2$.  To include some perpendicular links we
consider the configuration in Fig.\ \ref{fig-three-rows}a, where the
perpendicular links have been deleted at every third row only. The
approach is then to integrate out the upper and lower rows of spins in
each of these triple rows to give $L/3$ one-dimensional rows,
cf.~Fig.\ \ref{fig-three-rows}b.  These integrations may be done
analytically or by Monte Carlo simulations on that geometry.  The
integrations also give correlations to the neighboring and next
neighboring links, $\langle\phi\phi'\rangle$ and
$\langle\phi\phi''\rangle$ (cf.\ Fig.\ \ref{fig-prim-bis}), which
together give an effective variance $\sigma^2_{\mathrm eff}$:
\begin{equation}
  \sigma^2_{\mathrm eff}
  = \frac{1}{L}\left<\left(\sum_{i=1}^L \phi_i\right)^2\right>
  \approx \sigma^2 + \langle\phi\phi'\rangle + \langle\phi\phi''\rangle.
  \label{sigma-eff}
\end{equation}
With $L/3$ rows, as in Fig.~\ref{fig-three-rows}b, the expression for
the helicity modulus becomes $\Upsilon = T / (3\sigma^2_{\mathrm
  eff})$.  From our MC simulations on the geometry of
Fig.~\ref{fig-three-rows}a we get $\sigma^2 = 0.672$,
$\langle\phi\phi'\rangle = -0.120$, and $\langle\phi\phi''\rangle =
-0.026$. The first two of these numbers are easily obtained by
integrating analytically with symbolic software.  With \Eq{sigma-eff}
this gives $\sigma^2_{\mathrm eff} = 0.526$ and $\Upsilon/T = 0.633$
which is less than 20\% off $\Upsilon/T = 0.789$ obtained above.

To conclude we have presented evidence from simulations that the 2D
step model actually undergoes a KT transition. We have argued that the
reason for the stability of the low temperature phase against the
formation of free vortices is the $\ln L$-dependence of the free
energy for a vortex at a fixed position. From these results we are led
to the conclusion that the harmonic spin interaction is not a
necessary condition for a KT transition in a 2D spin model, and that
the KT transition is a more general phenomenon than has so far been
recognized.

The authors thank Prof.\ P. Minnhagen and Prof.\ S. Teitel for
critical reading of the manuscript. Financial support from the Swedish
Natural Science Research Council through Contract No.\ E-EG 10376-312
is gratefully acknowledged.


\begin{thebibliography}{10}
\bibitem{angle129}
This odd number comes from the desire to have a symmetric $U(\phi)$, together
  with one of the discrete $\phi$ being equal to zero.
\end{thebibliography}
\end{document}